\documentclass[twocolumn,showkeys,prd,nofootinbib,floatfix]{revtex4-1}

\usepackage{amsfonts,amsmath,amssymb}
\usepackage{graphicx,graphics}

\begin{document}

\title{\textbf{CAN FERMIONIC DARK MATTER MIMIC SUPERMASSIVE BLACK HOLES?}}\footnote{\textit{Essay written for the Gravity Research Foundation 2019 awards for essays on Gravitation}}

\author{C. R. Arg{\"u}elles,$^{1,*}$ A. Krut,$^{2}$ J.~A.~Rueda,$^{2,3}$ R.~Ruffini$^{2,4}$}

\affiliation{$^1$Instituto de Astrofisica de La Plata, (CCT La Plata, CONICET, UNLP), Paseo del Bosque, B1900FWA La Plata, Argentina}
\address{$^*$charly@carina.fcaglp.unlp.edu.ar (corresponding author)}
%\affiliation{$^1$Dipartimento di Fisica and ICRA, Sapienza Universit\`a di Roma, P.le Aldo Moro 5, I--00185 Rome, Italy}
%\affiliation{$^2$University of Nice-Sophia Antipolis, 28 Av. de Valrose, 06103 Nice Cedex 2, France}
\affiliation{$^2$ICRANet, Piazza della Repubblica 10, I--65122 Pescara, Italy}
\affiliation{$^3$INAF, Istituto de Astrofisica e Planetologia Spaziali, Via Fosso del Cavaliere 100, 00133 Rome, Italy}
\affiliation{$^4$INAF, Viale del Parco Mellini 84, 00136 Rome  Italy}
\address{andreas.krut@icranet.org, jorge.rueda@icra.it, ruffini@icra.it}
%\affiliation{$^5$ZARM, University of Bremen, 28359 Bremen, Germany}
%\affiliation{$^6$ICRANet-Rio, CBPF, Rua Dr. Xavier Sigaud 150, Rio de Janeiro, RJ, 22290--180, Brazil}

\date{\today}

\begin{abstract}

We analyze the intriguing possibility to explain both dark mass components in a galaxy: the dark matter (DM) halo and the supermassive dark compact object lying at the center, by a unified approach in terms of a quasi-relaxed system of massive, neutral fermions in general relativity. The solutions to the mass distribution of such a model that fulfill realistic halo boundary conditions inferred from observations, develop a highly-density core supported by the fermion degeneracy pressure able to mimic massive black holes at the center of galaxies. Remarkably, these \textit{dense core-diluted halo} configurations can explain the dynamics of the closest stars around Milky Way's center (SgrA*) all the way to the halo rotation curve, without spoiling the baryonic bulge-disk components, for a narrow particle mass range $mc^2 \sim 10$-$10^2$~keV.

\end{abstract}
% This fermion core can mimic either an intermediate-mass black hole (for dwarf galaxies), or a supermassive black hole (for spiral and elliptical galaxies).
% This result agrees with completely \textit{independent} lower DM mass bounds from cosmological/astrophysical origin (e.g.~phase-space bounds or Lyman-$\alpha$ forest), but has the advantage of solely relying on local Universe observations, regardless of the cosmological history of the particles. Further consequences of this theory about the formation of the most massive black holes at the center of quasars, or the strong lensing properties around SgrA*, are outlined.
\pacs{Valid PACS appear here}% PACS, the Physics and Astronomy
                             % Classification Scheme.

                              %display desired
\maketitle

%%%%%%%%%%%%%%%%%%%%%%%%%%%%%%%%%%%%%%%%%%%%%%%%%%%%%%%%%
%%%%%%%%%%%%%%%%%%%%%%%%%%%%%%%%%%%%%%%%%%%%%%%%%%%%%%%%%
\section{Introduction}
%%%%%%%%%%%%%%%%%%%%%%%%%%%%%%%%%%%%%%%%%%%%%%%%%%%%%%%%%
%%%%%%%%%%%%%%%%%%%%%%%%%%%%%%%%%%%%%%%%%%%%%%%%%%%%%%%%%

%-1st part) General (didactical) basic intro + including some main references to historical works - from icranet journal summary +.\\

Since about a century, astronomers and astrophysicists have gathered and analyzed data coming from galaxies, either small, or large ellipticals, or clumped in large clusters. This have revealed that about the $85\%$ of the matter content of the Universe cannot be made of any of the building blocks we know (such as electrons, protons, neutrons, or its combinations). They came to the conclusion that the gravity exerted by all these possible known forms of matter, as combined in stars, gas or dust, is not enough to explain the observed stability and the kinematic properties in galaxies \cite{edbk1}: an extra matter content was needed, called dark matter (DM).

A consensus has been reached within the scientific community about the generic nature of the DM, pointing towards an unknown fundamental massive particle created at the dawn of times. At certain moment in the evolution and cooling down of the Universe (prior to recombination epoch), these particles started to gather together by its own self-gravity into many different clumps of matter, called DM halos. Such pristine agglomerations are spherical configurations which constitute the progenitors of the galaxies we see today. After complex merging histories of such DM seeds, galactic structures in the local Universe end up enveloped in quasi-virialized dark halo components, spreading typically about ten times the extension of the bright and normal matter, as in the case of our Milky Way.

An important open question in this field is precisely how the DM is distributed throughout a galaxy, as well as the exact nature and mass of the DM constituent particle. The traditional approach to tackle this issue is given in terms of big numerical simulations involving a large amount (up to billions) of classical point masses for adequate initial conditions within the cold dark matter (CDM) paradigm \cite{jpap1}. While such simulations provide the needed dark mass to account either for the observed large scale structure of the Universe (above $\sim 100$~Mpc) down to the outer rotation curve of a galaxy, it has several problems on smaller scales below $10$~kpc \cite{jpap2}.

Alternative approaches to describing quasi-relaxed DM halos in terms of fundamental particles (i.e. allowing to consider the quantum nature of the DM constituents), are being recently considered since they may provide solutions to many of the unsuccessful predictions of the CDM cosmology on short scales (see e.g. \cite{jpap3} for the case of bosons, and \cite{jpap4} for the case of fermions). It is the purpose of this essay to show the main astrophysical consequences that arise by considering a self-gravitating system of massive fermions, distributed in phase-space by a Fermi-Dirac distribution function with a cutoff in momentum space (\ref{fcDF}), under the assumption of thermodynamic equilibrium in general relativity (GR). It is important to mention that such a quantum-statistical phase-space is not given \emph{ad-hoc}, but it can be obtained as a (quasi) stationary solution of a generalized thermodynamic Fokker-Planck equation for fermions, including the physics of collisionless (violent) relaxation and evaporation \cite{jpap5} appropriate to deal with non-linear structure formation. Indeed, such phase-space solutions are there shown to fulfill a maximization (coarse-grained) entropy principle (second law of thermodynamics) during the complex (collisionless) relaxation process, until the halo enters in the steady state we observe.

%%%%%%%%%%%%%%%%%%%%%%%%%%%%%%%%%%%%%%%%%%%%%%%%%%%%%%%%%
%%%%%%%%%%%%%%%%%%%%%%%%%%%%%%%%%%%%%%%%%%%%%%%%%%%%%%%%%
\section{Theoretical framework: the Ruffini-Arg\"uelles-Rueda (RAR) model}
%%%%%%%%%%%%%%%%%%%%%%%%%%%%%%%%%%%%%%%%%%%%%%%%%%%%%%%%%
%%%%%%%%%%%%%%%%%%%%%%%%%%%%%%%%%%%%%%%%%%%%%%%%%%%%%%%%%

The RAR model consists in a self-gravitating system of massive fermions (spin $1/2$) in hydrostatic and thermodynamic equilibrium within GR. More specifically, we work in the hydrodynamic approximation, and solve the Tolman-Oppenheimer-Volkoff (TOV) equations for a perfect fluid whose equation of state (EOS) takes into account (i) the relativistic effects of the fermionic constituents, (ii) finite temperature effects, and (iii) the escape of particle effects at large momentum ($p$) through a cut-off in the Fermi-Dirac distribution $f_c$ given in eq.~(\ref{fcDF}). Such a model was first presented in \cite{jpap4} accounting for (i) and (ii), and then extended in \cite{jpap6,jpap7} including for (iii) as well. Moreover it is the morel general of its kind, given it does not work under the fully-Fermi-degeneracy approximation, nor in the diluted-Fermi regime.
\begin{equation}
f_c(\epsilon\leq\epsilon_c) = \frac{1-e^{(\epsilon-\epsilon_c)/kT}}{e^{(\epsilon-\mu)/kT}+1}, \qquad f_c(\epsilon>\epsilon_c)=
0\, ,
\label{fcDF}
\end{equation}
where $\epsilon=\sqrt{c^2 p^2+m^2 c^4}-mc^2$ is the particle kinetic energy, $\mu$ is the chemical potential with the particle rest-energy subtracted off, $T$ is the temperature, $k$ is the Boltzmann constant, $c$ is the speed of light, and $m$ is the fermion mass. The full set of dimensionless-parameters of the model are defined by the temperature, degeneracy and cutoff parameters, $\beta=k T/(m c^2)$, $\theta=\mu/(k T)$ and $W=\epsilon_c/(k T)$, respectively.

The corresponding 4-parametric fermionic EOS (at given radius $r$): $\rho(\beta,\theta,W,m), P(\beta,\theta,W,m)$, where $\rho$ and $P$ are the mass-energy density and pressure, is directly obtained as the corresponding integrals (bounded from above by $\epsilon \leq \epsilon_c$)  over momentum space of $f_c(p)$ (explicited in \cite{jpap6}). The stress-energy tensor is the one of a perfect fluid with the above density and pressure. 
%Such components of the EOS correspond to the diagonal part of the stress-energy tensor in the Einstein equations, which are solved under the perfect fluid approximation (i.e. TOV as stated above) within 
The background metric with spherical symmetry is ${\rm d}s^2 = e^{\nu}c^2 {\rm d}t^2 -e^{\lambda}{\rm d}r^2 -r^2 {\rm d}\Theta^2 -r^2\sin^2\Theta {\rm d}\phi^2$, where ($r$,$\Theta$,$\phi$) are the spherical coordinates, and $\nu$ and $\lambda$ depend only on the radial coordinate $r$. The Einstein equations are solved together with the Tolman and Klein thermodynamic equilibrium conditions (i.e. zero and first law of thermodynamics in GR) conforming a coupled system of integro-differential equations as detailed in \cite{jpap6}.

The more general solutions to the RAR model equations develop a \textit{dense core-diluted halo} behavior (see the profiles for the Milky Way, as well as for generic dwarfs, spirals and ellipticals in Figs.~\ref{fig:rhotot-MW}, \ref{fig:vrot-MW} and \ref{fig:profiles}) which is explicitly manifest in the density profile and velocity rotation curve (the later calculated from the GR expression of the circular velocity of a test particle in the above metric) as follows:
\begin{itemize}
\item
an inner core with radius $r_c$ (defined at the first maximum of the rotation curve) of almost constant density governed by Fermi degeneracy where general relativistic effects become appreciable (see the high positive values of the degeneracy parameter responsible for this in upper panel of Fig.~2 in \cite{jpap6});
\item
an intermediate region with a sharply decreasing density distribution where quantum corrections are still important (see the transition from positive to negative values of the degeneracy parameter in upper panel of Fig.~2 in \cite{jpap6}), followed by an extended and diluted plateau, until the RAR halo scale-length $r_h$ is achieved (defined at the second maximum of the rotation curve); and
\item
a Boltzmannian density tail with negligible general relativistic effects (see the highly negative values of the degeneracy parameter in the upper panel of Fig.~2 in \cite{jpap6}) showing a behavior $\rho\propto r^{-n}$ with $n>2$ due to the cutoff constraint, implying a DM halo naturally bounded in radius (i.e. when $\rho\approx 0$ occurring when the escape energy $\epsilon_c\propto W$ approaches $0$), as can be seen in Fig.~\ref{fig:rhotot-MW}, once applied the Milky Way phenomenology.
\end{itemize}

%%%%%%%%%%%%%%%%%%%%%%%%%%%%%%%%%%%%%%%%%%%%%%%%%%%%%%%%%
%%%%%%%%%%%%%%%%%%%%%%%%%%%%%%%%%%%%%%%%%%%%%%%%%%%%%%%%%
\section{Astrophysical applications}
%%%%%%%%%%%%%%%%%%%%%%%%%%%%%%%%%%%%%%%%%%%%%%%%%%%%%%%%%
%%%%%%%%%%%%%%%%%%%%%%%%%%%%%%%%%%%%%%%%%%%%%%%%%%%%%%%%%

%\emph{The Milky Way}\\
\subsection*{The Milky Way}

The phenomenology of the RAR model when applied to our Galaxy, is implemented as follows: we adopt as boundary conditions for the solution to the RAR equations: (i) a DM halo mass with the observed value at two different radial locations in the Galaxy: a DM halo mass $M_{\rm DM}(r=40~{\rm kpc}) = 2\times 10^{11} M_\odot$, consistent with the dynamics of the outer DM halo, as constrained using the Galactic Sagittarius (Sgr) dwarf satellite in \cite{jpap8}, and $M_{\rm DM}(r=12~{\rm kpc}) = 5\times 10^{10} M_\odot$, as constrained in \cite{jpap9}. Simultaneously, we require a RAR dense-core of mass $M_{\rm DM}(r=r_c)\equiv M_c= 4.2\times10^6 M_\odot$ enclosed \textit{within} the radius $r_{p(S2)}=6\times 10^{-4}$~pc (e.g.~$r_c\leq r_{p(S2)}$), the value of the S2 star pericenter as observed in \cite{jpap10}. This last condition makes explicit the aim of searching for an alternative to the supermassive black hole (BH) scenario. Such a phenomenology implies three boundary conditions for the three free RAR model central parameters, once the particle mass is given.

The family set of RAR density profiles and rotation velocity curves ($v_{\rm DM}$) which fulfills these conditions are shown in Figs.~\ref{fig:rhotot-MW} and \ref{fig:vrot-MW}, respectively, where the excellent fit to the total (observed) rotation curve $v_{\rm rot}^2(r)=v_{\rm b}^2(r)+v_{\rm d}^2(r)+v_{\rm DM}^2(r)$ (shown in red, including for the bulge $v_{\rm b}$ and disk $v_{\rm d}$ baryonic components in green as defined in \cite{jpap6}) is given, from $10^{-7}$~pc all the way to $10^5$~pc. Notice we have used the extended high resolution full rotation curve data $v_{\rm rot}$ of the Milky Way as obtained in \cite{jpap9} from the inner bulge up to the halo scales (in blue dots including for error bars). See Fig.~\ref{fig:RARNFWzoom-vel} as well for a zoom on outer halo scales comparing with the phenomenological Navarro-Frenk-White (NFW) profile. The main outcome regarding the particle mass constraints can be summarized as follows (see \cite{jpap6} for further details):

\begin{figure}%
	\centering%
	\includegraphics[width=\hsize]{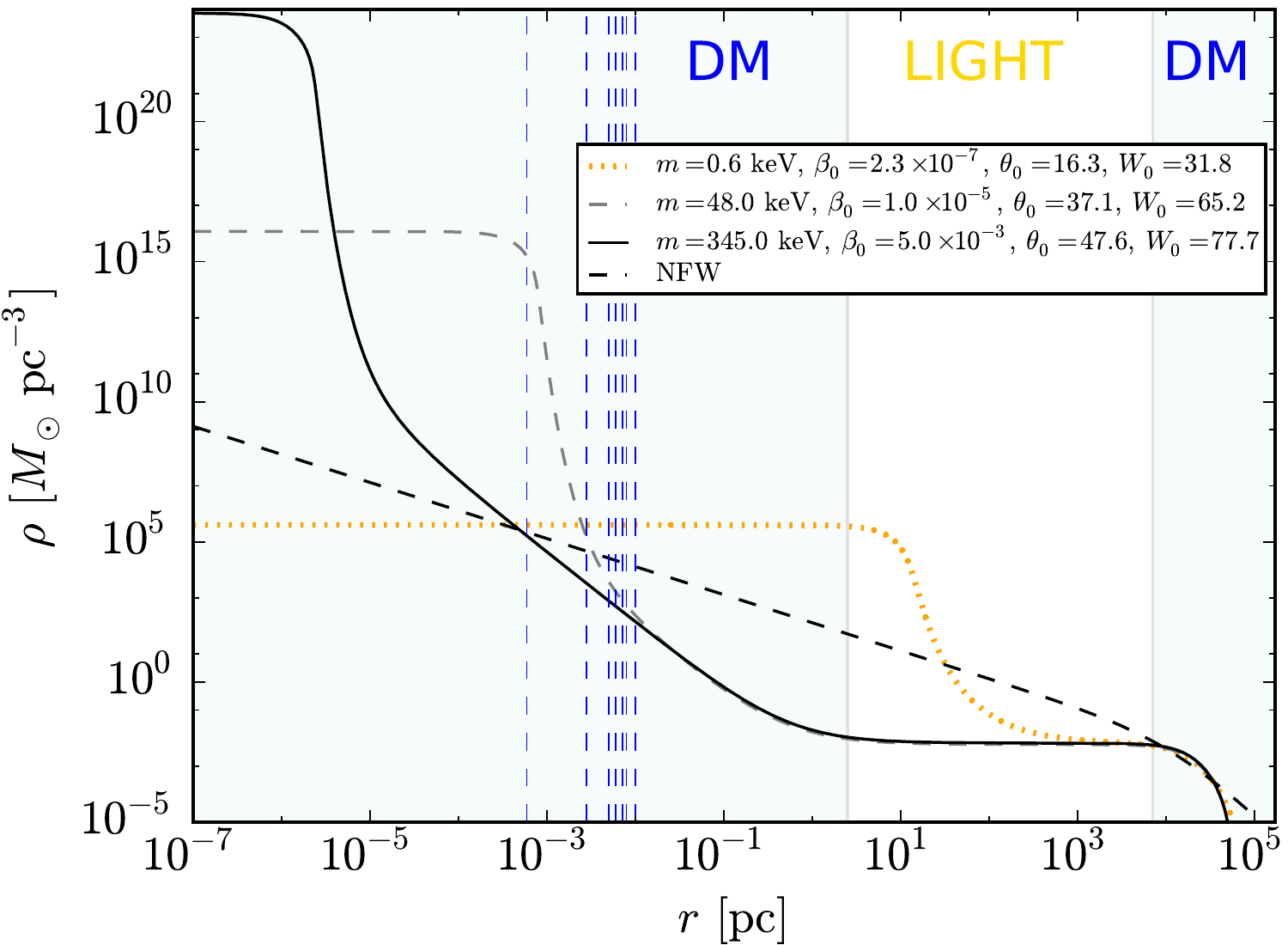}
	\caption{(Color online) Theoretical density profiles from $10^{-7}$~pc all the way to $10^5$~pc, for three representative fermion masses in the $mc^2\sim$keV region: 0.6~keV (dotted yellow curve), 48~keV (long-dashed gray curve) and 345~keV (solid black curve). The dashed-blue lines indicate the position of the S-cluster stars. We show for the sake of comparison the NFW density profile (dashed-black curve). See the text for additional details.}
	\label{fig:rhotot-MW}
\end{figure}

%
% Is here the enumeration really important? Maybe a simple/neutral itemize will do it better.
\begin{itemize}
\item
For fermion masses $mc^2 = 48$--$345$~keV, the RAR solutions with corresponding initial parameters ($\beta_0$, $\theta_0$, $W_0$) explain the Galactic DM halo and, at the same time, provide an alternative to the central BH scenario. The mass lower bound in $m$ is imposed by the dynamics of the stellar S-cluster. Namely, the quantum core radius of the solutions for $mc^2 \geq 48$~keV are always smaller or equal than the radius of the S2 star pericenter, i.e. $r_c \leq r_{p(S2)} = 6\times 10^{-4}$~pc~$\approx 1.5\times 10^3 r_{\rm Sch}$. With $r_{\rm Sch}$ the Schwarzschild radius of a $4.2\times10^6 M_\odot$ BH.
\item
There is a mass upper bound of $mc^2 = 345$~keV that corresponds to the last stable configuration before reaching the critical mass for gravitational collapse ($M_c^{\rm cr}\propto m_{\rm Planck}^3/m^2$), which is calculated following the turning point criterion for core-collapse in \cite{jpap11}. The core radius of the critical configuration is $r_c\approx 4 \,r_{\rm Sch}$.
\item
The fermion mass range $mc^2 \lesssim 10$~keV is firmly ruled out by the present analysis because the corresponding rotation curve exceeds the total velocity observed in the baryonic (bulge) dominated region $r\approx 2$--$100$~pc (including upper bound in error bars). We plot for example the highly exceeding case of $mc^2 = 0.6$~keV (dotted yellow curve) in Fig.~\ref{fig:vrot-MW}, where the DM distribution produces a large overshoot over the observed rotation curve.
\item
There is an intermediate fermion mass range $mc^2 = 10$-- $48$~keV where the theoretical rotation curve is not in conflict with any of the observed data and DM inferences given in \cite{jpap9}, but the compactness of the quantum core is not enough to be an alternative to the central BH scenario for SgrA*.
\item 
In the case of the more compact RAR cores alternative to the SgrA* BH  (e.g. $mc^2 \sim 100$~keV), a possible observational scenario able to distinguish between the two is through lensing techniques: the DM cores do not show a photon sphere implying that they do not cast a shadow (they are transparent), opposite to the BH case. Instead the DM compact cores can generate multiple images and Einstein rings on the very central regions (at tens to hundreds Schwarzschild radius) as demonstrated in \cite{jpap12} by using strong lensing calculations. Future observations (likely needing larger angular resolution) by the Event Horizon Telescope (EHT) may help to discriminate between the two models.
\end{itemize}

\begin{figure}%
	\centering%
	\includegraphics[width=\hsize]{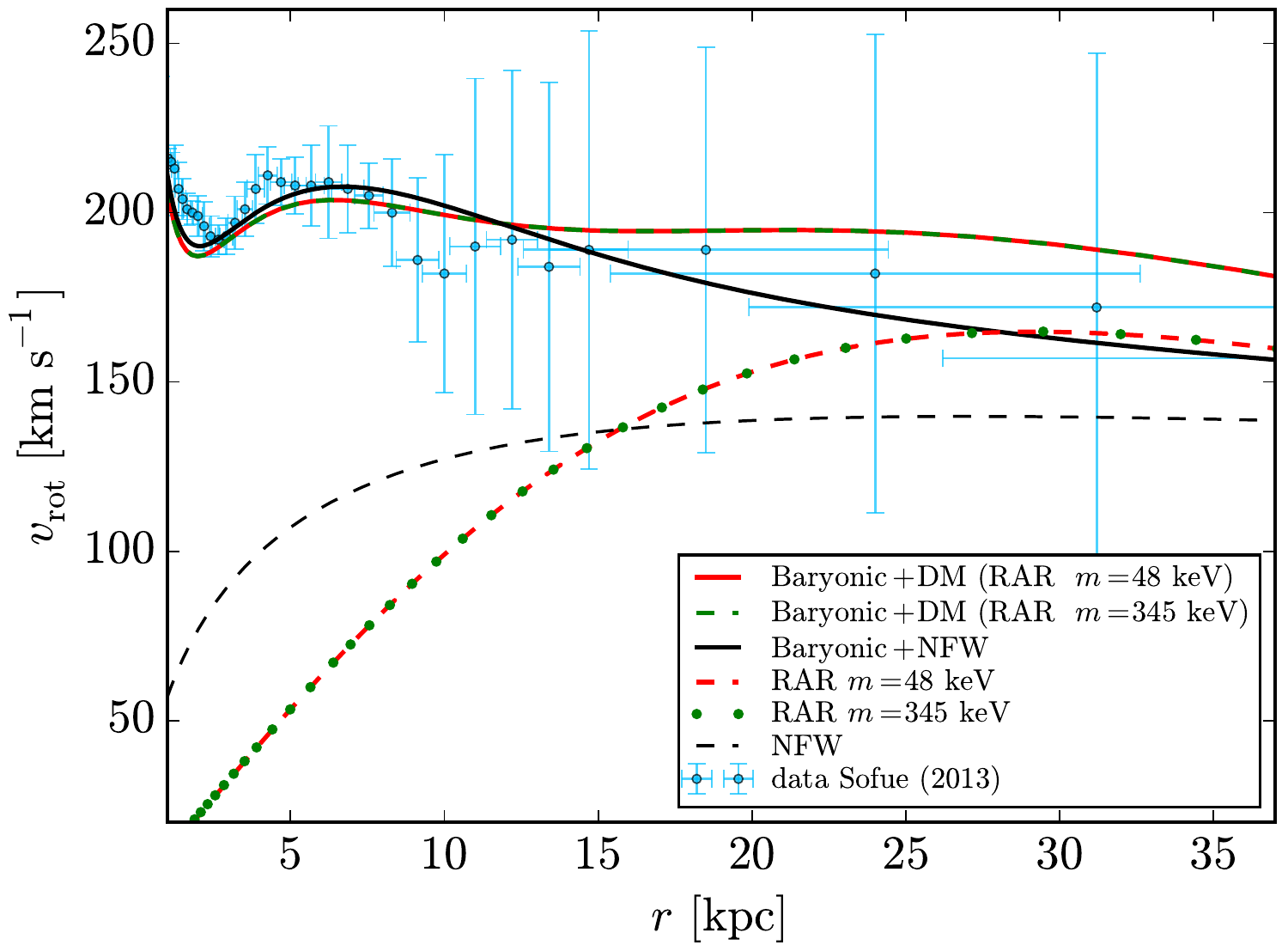}
	\caption{(Color online) Comparison of the total (DM plus baryonic) rotation curves as given by the RAR model (for a fermion mass $m c^2=48$~keV and 345~keV) and the phenomenological NFW mode in the region $r = 1$--$35$~kpc. Within the region $1$ -- several~kpc, the contribution of the baryonic components dominates respect to the DM, while at distances $r\gtrsim 10$~kpc there is, instead, an increasing dominance of the DM component.}
	\label{fig:RARNFWzoom-vel}
\end{figure}

\begin{figure}%
	\centering%
	\includegraphics[width=\hsize]{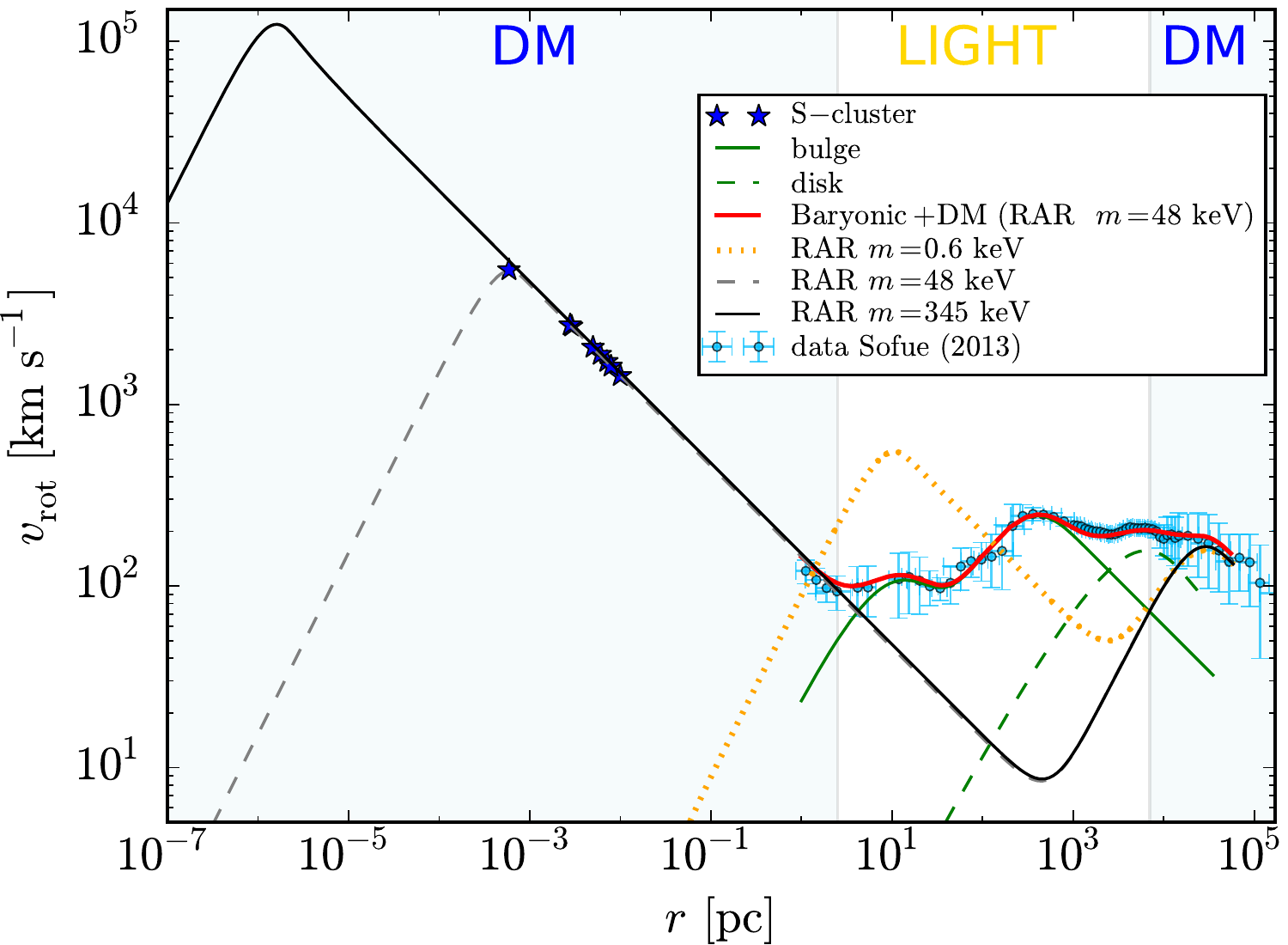}
	\caption{(Color online) Theoretical RAR rotation curves from $10^{-7}$~pc all the way to $10^5$~pc, for three representative fermion masses in the $mc^2\sim$keV region: 0.6~keV (dotted yellow curve), 48~keV (long-dashed-gray curve) and 345~keV (solid-black curve). These RAR solutions are in agreement with all the Milky Way observables from $\sim 10^{-3}$~pc to $\sim 10^{5}$~pc. For the case of $mc^2=48$~keV, we include the total rotation curve (red-thick curve) including the total baryonic (bulge + disk) component. The star symbols represent the eight best resolved S-cluster stars. See the text for additional details.}
	\label{fig:vrot-MW}
\end{figure}

%\emph{From dwarf to elliptical galaxies}\\
\subsection*{From dwarf to elliptical galaxies}

We now choose three samples of different galaxy types covering: (i) well resolved dwarf spheroidal (dSph) satellites of the Milky Way \cite{jpap13}; (ii) nearby (high resolution) disk galaxies from the THINGS data sample \cite{jpap14}; and (iii) elliptical galaxies analyzed through weak lensing as studied in \cite{jpap15}. The observational inferences of the DM content for such (far away) galaxy types are limited to a narrow window of galaxy radii, usually lying just above the baryonic dominance region. Thus, we adopt in this case a similar methodology compared to the Milky Way analysis shown above, but limited to radial halo extents where observational data is available. In particular we will select as the only boundary conditions (as allowed by the DM halo observational inferences (i)-(iii) above), a characteristic halo radius $r_h$ located at the maximum of the halo rotation curve (with corresponding halo mass $M_h \equiv M_{\rm DM}(r_h)$), obtained from an averaging procedure for each galaxy-type sample: $(r_h,M_h)_{\rm d}=(400 \rm pc, 3\times10^7 M_\odot); (r_h,M_h)_{\rm s}=(50 \rm kpc, 10^{12} M_\odot), (r_h,M_h)_{\rm e}=(90 \rm kpc, 5\times10^{12} M_\odot)$ for dSphs, spirals, and elliptical galaxies, respectively.

\begin{figure*}%
	\centering%
	\includegraphics[width=\hsize]{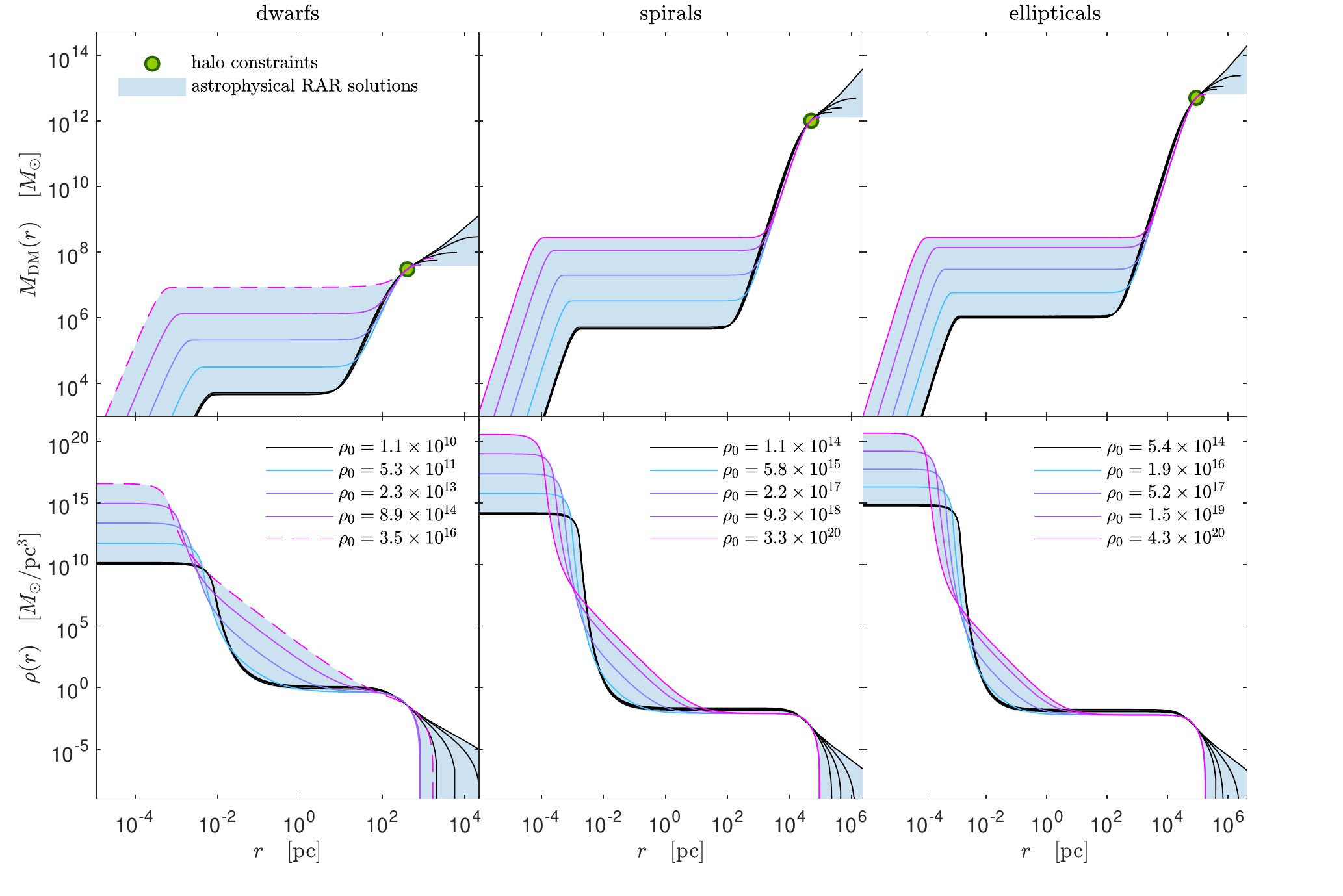}
	\caption{Astrophysical RAR solutions, for the relevant case of $mc^2 = 48$~keV, fulfill observationally given DM halo restrictions $(R_h,M_h)$ for typical dwarf (left), spiral (middle) and elliptical galaxies (right). Shown are density profiles (bottom), and DM mass distributions (top). The full window for each galaxy type is illustrated by a blue shaded region and enveloped approx. by $5$ benchmark solutions inside. Each solution is labeled with the central density in units of $M_\odot$~pc$^{-3}$. The continuous-magenta curves, occurring only for spiral and elliptical galaxies, indicates the critical solutions which develop compact critical cores (before collapsing to a BH) of $M_c^{\rm cr}=2.2\times10^8 M_\odot$. The dashed-magenta curves for dwarfs are limited (instead) by the astrophysical necessity of a maximum in the halo rotation curve. The bounding black solutions correspond to the ones having the minimum core mass (or minimum $\rho_0$) which in turn imply larger cutoff parameters. Thus, these solutions develop more extended density tails, where $\rho\propto r^{-2}$ is the limiting isothermal density tail achieved when $W_0 \rightarrow\infty$.}
	\label{fig:profiles}
\end{figure*}

We thus systematically calculate, for the relevant case of $mc^2=48$~keV motivated by the Milky Way analysis, the set of astrophysical RAR mass and density profiles, by covering the full free parameter-space ($\beta_0,\theta_0,W_0$)  matching the halo constraints ($r_h,M_h$) for each galaxy type. There is a relatively narrow set of solutions fulfilling such bounds, illustrated as a blue-shaded region in Fig.~\ref{fig:profiles} and enveloping the five benchmark (example) solutions labeled through their central densities. They encompass a window of possible core and total DM masses $M_c$ and $M_{\rm tot}$ (the later defined at the boundary of the DM configuration) for each galaxy structure, which are direct \textit{predictions} of the RAR theory. Interestingly, for an unique particle mass $mc^2=48$~keV, there is the possibility among the RAR core masses (concentrated below milliparsec scales) to have $M_c\sim 10^4$--$10^6 M_\odot$ for dSphs, and $M_c\sim 10^5$--$10^8 M_\odot$, for typical spiral and elliptical, thus providing natural alternatives to intermediate and supermassive BHs, respectively (see \cite{jpap7} for further details). The limiting value of $2.2\times10^8 M_\odot$ (for $mc^2=48$~keV) corresponds to the critical core mass above which the gravitational collapse towards a supermassive BH takes place, and remarkably matches the mass value of the supermassive BHs observed.

The most relevant consequences of these RAR profile predictions relies in the fact they are in good agreement with the observational galaxy scaling relations: (1) the Ferrarese relation indicating that the larger the DM halo mass, the larger the supermassive dark central object \cite{jpap16}; and (2) the nearly constant DM surface density of galaxies \cite{jpap15} (the later $\propto \rho(r_{pl})\,r_h$, with $\rho(r_{pl})$ the density at plateau), as demonstrated in \cite{jpap7}. Finally, supermassive dark compact objects of $10^9 M_\odot$ or larger (as detected in the largest elliptical galaxies) are usually associated with active galaxies and are characterized by a clear X-ray and radio emissions as well as jets. If such objects are indeed BHs, it may be explained starting from a BH seed of mass $\sim 10^8 M_\odot$ formed out of the collapse of our critical DM cores. After its formation, such a BH seed might start a baryonic and/or DM accretion process from their massive galactic environment ($M_g\sim 10^{12} M_\odot$). Indeed an accretion of $\sim 1 \%$ of the (inner) mass of the galaxy onto the critical core (over cosmological time-scales), would be enough to explain the formation of the largest supermassive BH masses without the need to go for unrealistic accretion rates.

As a final remark, we would like to mention an important aspect of the particle mass range of few $10$ -- $100$~keV obtained here solely from local Universe observables. Namely, it produces practically the same behavior in the power spectrum (down to Mpc scales) as that of standard $\Lambda$CDM cosmologies, thus providing the expected large-scale structure features. In addition, it is not `too warm' (i.e. our masses are larger than $mc^2\sim 1$ -- $3$~keV) to enter in tension with current Lyman-$\alpha$ forest constraints and the number of Milky Way satellites, as in standard $\Lambda$WDM cosmologies.

%%%%%%%%%%%%%%%%%%%%%%%%%%%%%%%%%%%%%%%%%%%%%%%%%%%%%%%%%%%%%%
%%									 References                         %%
%%%%%%%%%%%%%%%%%%%%%%%%%%%%%%%%%%%%%%%%%%%%%%%%%%%%%%%%%%%%%%

%\bibliographystyle{apsrev}
%\bibliography{biblio}

%%%%%%%%%%%%%%%%%%%%%%%%%%%%%%%%%%%%%%%%%%%%%%%%%%%%%%%%%%%%%%

\end{document}